\documentclass[twocolumn,showpacs,amsmath,amssymb,prl]{revtex4}

\usepackage{epsfig}

\begin{document}

\title{Transient dynamics of the
nonequilibrium Majorana resonant level model}
\author{A.~Komnik}
\affiliation{Institut f\"ur Theoretische Physik, Universit\"at
Heidelberg,
 Philosophenweg 19, D-69120 Heidelberg, Germany}
\date{\today}

\begin{abstract}
The Majorana resonant level model (MRLM) describes the
universality class of the two-channel/terminal Kondo model at the
Toulouse point as well as of a resonant level between two
half-infinite Tomonaga--Luttinger liquids. We analyze the time
evolution of the electric current and of the population function
after an instantaneous switching on of the tunneling coupling. We
find that the only timescale, which governs the relaxation of the
initial dot preparation is the inverse contact transparency
$\Gamma$, whatever the dot offset energy $\Delta$, applied bias
voltage or temperature. The voltage alone determines the
superimposed oscillatory behavior of the observables for weak
detuning $|\Delta|<\Gamma/2$. In the opposite case of strong
detuning $|\Delta|>\Gamma/2$ a beating pattern emerges. For the
current the finite temperature plays the similar role as the
hybridization. The dot population function dynamics approaches
that of a resonant ($\Delta=0$) setup upon increasing the voltage
or/and temperature.
\end{abstract}

\pacs{72.10.Fk, 73.63.-b, 71.10.Pm}

\maketitle
%\section{Introduction}
During the last decades we witnessed a soaring interest in the
nonequilibrium properties of quantum impurity models. It has been
motivated by the advances in manufacturing of semiconductor based
nanostructures, in which only a small number of electronic levels
participate in the charge and spin transport. The experimentalists
not only succeeded in observing the Kondo effect in such
structures, but also were able to demonstrate phonon-assisted
charge transport through single molecules, see
e.~g.~\cite{goldhaber-gordon}.
%[\onlinecite{goldhaber-gordon,kouwenhoven,weisvonklitzing,ParkMcEuen}].
From the theoretical point of view such zero-dimensional systems
can be seen as realizations of quantum impurity models with or
without internal degrees of freedom. Typical examples are the
Anderson model, Tomonaga--Luttinger liquids (TLLs) and fractional
quantum Hall edge states with impurities, resonant levels coupled
to a Holstein phonon etc. Many of their equilibrium properties can
be calculated with the help of by now quite substantial number of
techniques. On the contrary, in the nonequilibrium one is
especially missing the extremely powerful Bethe ansatz and
integrability methods. Although their adaptation to setups far
from equilibrium has already begun, a number of open issues are
yet to be settled.\cite{KSLPRL}
%\cite{KSLPRL,mehta:216802,boulat:140601,doyon:076806}

Fortunately, at least in the universality class of Kondo models
there are some exactly solvable points which occur at nontrivial
constellations of system parameters: the Toulouse limits. Those
turn out to be the only fully featured testing grounds (apart from
numerical approaches, of course), which we have at our disposal in
the moment. In its canonical incarnation it is a special point in
the parameter space where a mapping of the conventional Kondo
model onto a resonant level model is possible.\cite{Toulouse,book}
For transport phenomena a two terminal topology is
necessary.\cite{EK} The original Hamiltonian is then mapped onto
that of a Majorana resonant level model (MRLM). It turns out to be
isomorph to the interacting resonant level in the
TLL.\cite{ourPRL} Hence, a solution of MRLM provides us with
predictions to a considerably broader class of situations, than
the original Kondo quantum dot. Thus far this kind of exact
solvability has been taken advantage of in a number of
contributions.\cite{SH,PhysRevLett.77.1821,komnik:216601} However,
only a fraction of them went beyond the steady state
calculations.\cite{PhysRevB.62.R16271,lobaskin:193303}
%\cite{PhysRevB.62.R16271,lobaskin:193303,AM}
The quantities of special interest are the transport current
evolution, several aspects of which has bee discussed in
Ref.~\cite{PhysRevB.62.R16271}, and the magnetization of the
impurity in the Kondo case, or the population of the resonant
level in the TLL resonant level setup. To the best of our
knowledge, the latter observables have not yet been addressed. We
would like to close this gap and present the full solution in the
case of instantaneous switching on of the tunneling hybridization.

%\section{The origin of the MRLM}

We first summarize the MRLM Hamiltonian in its most general form.
It features local Majorana fermions $a,b$ as well as four
different Majorana fermionic fields. They are organized into pairs
describing a \emph{principal} channel $\eta(x),\xi(x)$ and a
\emph{flavor} channel $\eta_f(x), \xi_f(x)$:
\begin{eqnarray}                        \label{MRLM1}
 H &=& H_0[\xi,\eta,\xi_f,\eta_f] - i \left[ \Delta \, a b +
 J_- \, b \, \xi_f(0) +
 J_+ \, a \, \eta_f(0)  \nonumber \right. \\
 &+& \left.
 \gamma_+ \, b \, \xi(0)
 + \gamma_- \, a \, \eta(0) \right] \, .
\end{eqnarray}
$H_0$ is responsible for the dynamics of the free (decoupled from
the local Majoranas) fields,
\begin{eqnarray}                      \label{H0prime}
 H_0 &=& i \int\, dx \, \Big[ \eta_f(x)
 \partial_x \eta_f(x) + \xi_f(x) \partial_x \xi_f(x)  \\ \nonumber
 &+&\eta(x) \partial_x \eta(x) + \xi(x) \partial_x \xi(x)
 + V \xi(x) \eta(x) \Big] \, .
\end{eqnarray}
$\Delta, J_\pm$ and $\gamma_\pm$ are different constant couplings
whereas $V$ is in a strict sense not a coupling but a chemical
potential and is related to the bias voltage applied in the
systems parental to MRLM. This is the reason why we employ
nonequilibrium diagrammatics in order to calculate the observables
of interest. These are the current
\begin{eqnarray}                     \label{current}
I = (i/2) \left[ \gamma_-  \, \langle a \, \xi(0) \rangle +
 \gamma_+ \, \langle b  \, \eta(0) \rangle \right]\, ,
\end{eqnarray}
and the dot occupation probability,
\begin{eqnarray}               \label{nd}
 n_d =
 \left(1 + i
 \, \langle a b \rangle \right)/2 \, .
\end{eqnarray}
The origin of the MRLM is twofold: it emerges in the (i) two
terminal Kondo model at the Toulouse point, and (ii) interacting
resonant level between two TLLs. In the situation (i) one starts
with the conventional Kondo Hamiltonian (we set $\hbar = v_F = e =
k_B = 1$), $ H = H_0 + H_J + H_M + H_V$, where, with
$\psi_{\alpha, \sigma}$ being the electron field operators in the
R,L (right/left) terminals,
\begin{eqnarray}\label{Hkondo}
 H_0 &=& i \sum_{\alpha=R,L} \sum_{\sigma=\uparrow,\downarrow}
 \int \, dx \,
 \psi^\dag_{\alpha \sigma}(x) \partial_x \psi_{\alpha \sigma}(x) \, ,
 \nonumber \\
 H_J &=& \sum_{\alpha, \beta = R,L} \sum_{\nu=x,y,z}
 J_\nu^{\alpha \beta}
 s^\nu_{\alpha \beta} \tau^\nu \, , \nonumber \\
 H_V &=& (V/2) \sum_\sigma \int \, dx \, ( \psi^\dag_{L \sigma}
 \psi_{L \sigma}
 - \psi^\dag_{R \sigma} \psi_{R \sigma}) \, , \nonumber \\
 H_M &=&
 -\mu_B g_i H \tau^z =
 - \Delta \tau^z \, .
\end{eqnarray}
Here $\tau^{\nu=x,y,z}$ are the Pauli matrices for the impurity
spin and ($\alpha,\beta=R,L$; $\sigma=\uparrow,\downarrow$;
$\sigma^\nu_{\sigma \sigma'}$ are the components of the $\nu$th
Pauli matrix)
\[
s^\nu_{\alpha \beta}=\sum_{\sigma,\sigma'}\, \psi_{\alpha
\sigma}^\dag(0) \, \sigma^\nu_{\sigma \sigma'} \, \psi_{\beta
\sigma'}(0) \, ,
\]
are the generalized electron spin densities in (or across) the
leads biased by a finite $V$. The last term in Eq.~(\ref{Hkondo})
stands for the magnetic field, $\Delta=\mu_B g_i H$. Following
Ref.~\cite{SH}, we assume $J_x^{\alpha \beta} = J_y^{\alpha \beta}
= J_\perp^{\alpha \beta}$, $J_z^{LL} = J_z^{RR} = J_z$ and
$J_z^{LR}=J_z^{RL}=0$. After bosonization, Emery-Kivelson
rotation, and refermionization (see details in \cite{SH} or
\cite{PhysRevB.68.235323}) and setting $J_z = 2 \pi$, one obtains
the  Toulouse point Hamiltonian (\ref{MRLM1}), where $J_\pm =
(J_\perp^{LL} \pm J_\perp^{RR})/\sqrt{2\pi a_0}$,
$\gamma_+=J_\perp = J_\perp^{RL}/\sqrt{2 \pi a_0}$, $\gamma_-=0$
($a_0$ is the lattice constant of the underlying lattice model)
and $a$ and $b$ being local Majorana operators originating from
the impurity spin $\boldsymbol{\tau}=(\tau^x,\tau^y,\tau^z)$:
${\bf \tau}^x = a$, ${\bf \tau}^y = b$. The fields $\eta_f$ and
$\xi_f$ in the spin--flavor sector are equilibrium (real) Majorana
fields, whereas $\eta$ and $\xi$ in the charge--flavor sector are
biased by the transport voltage $V$. The current through the
system is then given by (\ref{current}) and the impurity
magnetization by
\begin{eqnarray}
 m = \langle \tau^z \rangle = - i \langle \tau^x \, \tau^y \rangle
 = i \, \langle a \, b  \rangle =
 2 n_d - 1 \, .
\end{eqnarray}

The setup (ii) is the spinless TLL resonant level model with the
Hamiltonian
 $H = H_K + H_t + H_I$,
where $H_K$ is the kinetic part, $H_K = \Delta \, d^\dag d +
\sum_{i=R,L} H_0[\psi_i]$, describing the electrons in the leads
$H_0[\psi_i]$, and the resonant level with energy $\Delta$, the
corresponding electron operators being $d^\dag,d$. The dot can be
populated from either of the two leads ($i=R,L$) via electron
tunneling with amplitudes $\gamma_i$,
 $H_t = \sum_i  \gamma_i[ d^\dag \psi_i(0) + \mbox{h.c.}]$.
$H_I$ describes the electrostatic Coulomb interaction with the
strength $U$ between the leads and the dot,
 $H_I =  U \, d^\dag d \, \sum_i \, \psi_i^\dag(0) \psi_i(0)$.
The contacting electrodes are one-dimensional half-infinite
electron systems. We model them by chiral fermions living in an
infinite system.\cite{ourPRL} In the bosonic representation
$H_0[\psi_i]$ are diagonal even in presence of interactions (for a
recent review see e.g.~\cite{book}; we set the renormalized Fermi
velocity $v=v_F/g=1$, the bare velocity being $v_F$):
\begin{eqnarray}                 \label{Hi}
 H_0[\psi_i] = \frac{1}{4 \pi} \int \, dx \, [\partial_x \phi_i(x)]^2.
\end{eqnarray}
Here the phase fields $\phi_i(x)$ describe the slow varying
spatial component of the electron density (plasmons),
$\psi^\dag_i(x) \psi_i(x) = \partial_x \phi_i(x)/2 \pi \sqrt{g}$.
The electron field operator at the boundary is given by
 $\psi_i(0) = e^{i \phi_i(0)/\sqrt{g}}/\sqrt{2 \pi a_0}$.
 Here $g$ is the conventional TLL parameter related to the
bare interaction strength $U_{TLL}$ via $g=(1+U_{TLL}/\pi
v_F)^{-1/2}$.\cite{book,PhysRevB.46.15233} In the chiral
formulation the bias voltage amounts to a difference in the
densities of the incoming particles in both channels far away from
the constriction.\cite{eggergrabert} The current is then
proportional to the difference between the densities of incoming
and outgoing particles within each channel.

At generic $g \neq 1$ the problem cannot be solved exactly though.
However, at $g=1/2$ after a transformation of $d^\dag$ and $d$
operators to the spin representation of the form $\tau^x = (d^\dag
+ d)$, $\tau^y = - i (d^\dag - d)$, and $\tau^z =2 d^\dag d - 1$
one immediately observes that  the $U$-term is analogous to the
$\tau^z$--spin density coupling in the Kondo problem. Indeed, a
mapping onto the MRLM can be performed in the same way as in the
Kondo case and one again obtains Hamiltonian (\ref{MRLM1}) with
$\gamma_\pm = \gamma_L \pm \gamma_R$ and $J_\pm=0$.\cite{ourPRL}
As before we are interested in the transport current
(\ref{current}) and dot population (\ref{nd}). The time derivative
of $n_d$ is the displacement current in the system
$I_{\mbox{disp}}(t) = d n_d (t)/d t$. The currents through the
individual contacts $I_{L,R}$ can then be conveniently evaluated
using the relations\cite{schmidt:235110}
 $I_{L,R}(t) = I(t) \pm I_{\mbox{disp}}(t)/2$.
The effect of the asymmetry $\gamma_- \neq 0$ is similar to that
of a finite detuning $\Delta \neq 0$,\cite{PhysRevB.68.235323}
that is why in order to make the calculations and results more
lucid we concentrate on the latter case and set $\gamma_- =0$.
Then the current operators are identical in both (i) and (ii).

%\section{Calculation of the current}

In general the calculation of the observables is simplified by
their reduction to a related nonequilibrium Green's function (GF).
In the case of the transport current it is
  $G_{b \eta}(t,t') = - i \langle T_C \, b(t) \, \eta(t')
  \rangle$,
where $T_C$ is the time ordering along the Keldysh contour
$C$.\cite{PhysRevB.48.8487} Then the time dependent current is
given by
 $I(t) = - (\gamma/2) \, G_{b \eta} (t+0^+,t)$,
where by abuse of notation $G_{b \eta} (t,t')$ denotes the time
ordered GF and $\gamma = \gamma_+$. From now on we assume a
steplike switching on of the tunneling when $\gamma(t) = \gamma \,
\Theta(t)$, where $\Theta(t)$ is the Heaviside function. By an
expansion in $\gamma$ and resummation of the series one can show,
that the following reduction is valid:\cite{PhysRevB.48.8487}
\begin{eqnarray}
 I(t) = i \gamma \int_0^\infty d t' \left[ D_{bb}(t,t') \, g_{\xi
 \eta}(t'-t)
 \nonumber \right. \\ \left.
 - D^<_{bb}(t,t') \, g^>_{\xi \eta}(t'-t) \right] \, .
\end{eqnarray}
$D_{bb}(t,t') = - i \langle T_C b(t) \, b(t') \rangle$ is the
\emph{exact} homogeneous $b$-Majorana GF. $g_{\xi \eta}(t,t') = -
i \langle T_C \xi (t) \, \eta(t') \rangle$ are the zero order in
tunneling GFs. Remarkably, all its Keldysh components are equal
and given by
 $g_{\xi \eta}(\omega) = (n_L - n_R)/2$,
where $n_{R,L}$ are the Fermi distributions in the right(left)
electrode.\cite{PhysRevB.68.235323} In the Kondo setup we model
the electrodes by wide flat band with constant density of states
$\rho_0$. In the resonant level TLL setup $\rho_0$ is equal to the
energy-independent prefactor in the density of states in vicinity
of the Fermi edge. Here the applied voltage must be
doubled.\cite{AndersonFCS} Then for the current through the system
we obtain:
\begin{eqnarray}                      \label{currentfromft}
 I(t) =
  i \frac{\Gamma T}{2} \int_0^{\infty} d t' \, D^R_{bb}(t-t') \,
\frac{\sin[V (t'-t)]}{\sinh[\pi T (t'-t)]}
 \, ,
\end{eqnarray}
where $\Gamma = \rho_0 \gamma^2/2$. Thus everything is determined
by the retarded $D^R_{bb}$ only. In the time domain we obtain:
\begin{eqnarray}
  \label{1stequation}
 && D_{bb}^R (t,t') = D^{(0) R}_{bb}(t-t')
  \\ \nonumber
 &+& \int_0^\infty  d
 t_1 \, d t_2 \, D_{bb}^{(0) R}(t,t_1) \,
 \Sigma^{R}(t_1-t_2) \, D_{bb}^R(t_2, t') \, ,
\end{eqnarray}
where the self-energy is due to the tunneling only and is up to a
prefactor identical to the homogeneous $\eta$-GF. It is most
compact in the energy representation:
\begin{eqnarray}
 \Sigma^R(\omega) &=& - i \Gamma \, , \, \, \,
 \Sigma^<(\omega) = - i \Gamma (n_L + n_R) \, ,
 \nonumber \\
 \Sigma^>(\omega) &=& - i \Gamma (n_L + n_R -2) \, .
\end{eqnarray}
\begin{figure}[t]
  \centering
  \vspace*{0.3cm}
  \includegraphics[width = 0.40 \textwidth]{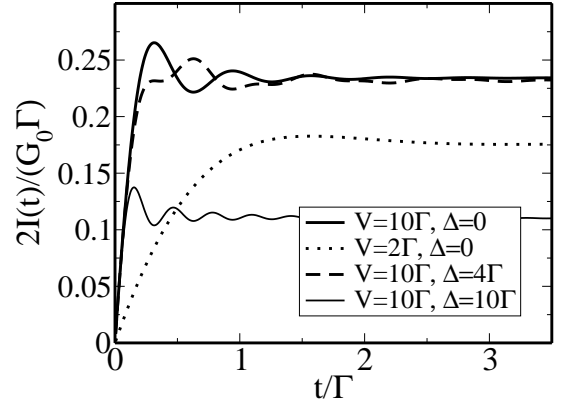}
  \caption{Full current through the constriction at zero temperature for
  different voltages $V$ and dot offset energies $\Delta$. The current is
  measured in units of $G_0 \Gamma/2$, where $G_0=2e^2/h$ is the conductance
  quantum.}
  \label{MajoranaSwitchingCurrentPlots}
\end{figure}
To make progress we use
\begin{eqnarray}                    \label{retintime}
 D^{(0) R}_{bb}(t,t') = - i \Theta(t - t') \, \cos[ \Delta (t-t')] \,
 ,
\end{eqnarray}
for the GF in the absence of coupling, which is valid at $J_\pm
=0$.\footnote{The calculation without this restriction is exactly
the same up to some minor adjustments. For the details see e.~g.
Appendix A of Ref.~\cite{AndersonFCS}.} It can be shown, that the
solution of (\ref{1stequation}) is translationally invariant in
time and has the structure $D_{bb}^R (t,t') = D_{bb}^R (t-t') = -
i \Theta(t-t') \, f(t-t')$, where
\begin{eqnarray}
  f(t) = f^{(0)}(t) - \Gamma \int_{0}^{t} d \tau \,
f^{(0)}(t- \tau) \, f(\tau) \, ,
\end{eqnarray}
with $f^{(0)}(t) = \cos(\Delta t)$. This is a Volterra integral
equation of the second kind solvable by the Laplace
transformation. We observe however, that the equation for the
retarded dot GF in the static case without switching effects has
exactly the same form.\footnote{This feature is shared by the
corresponding equation for the conventional resonant level even
beyond the wide flat band model.\cite{schmidt:235110}} The
explanation for that is the functional similarity of our switching
method and of the natural steplike time dependence of the retarded
GF. The solution in the energy domain is known to be given by
\begin{eqnarray}
 D_{bb}^R(\omega) = \frac{\omega}{\omega^2 - \Delta^2 + i \omega
 \Gamma} \, .
\end{eqnarray}
Transformed back it yields
\begin{eqnarray}
f(t) = \frac{e^{-\Gamma t/2}}{2 \Omega}
 \left[ 2 \Omega \, \cosh\left(t \Omega \right) - \Gamma \, \sinh
 \left( t \Omega \right) \right] \, ,
\end{eqnarray}
with $\Omega = \sqrt{(\Gamma/2)^2 - \Delta^2}$ for \emph{weak
detuning} $|\Delta| < \Gamma/2$. In case of \emph{strong
detuning}, when $|\Delta| > \Gamma/2$ the function is found by an
analytic continuation. Knowing that the retarded GF is indeed
translationally invariant in time allows for further
simplification of (\ref{currentfromft}):
\begin{eqnarray}
 I(t) = \frac{\Gamma T}{2} \int_0^t d \tau \, f(\tau) \,
 \frac{\sin(V \tau)}{\sinh(\pi T \tau)} \, .
\end{eqnarray}
In general the current shows up two constituents: the transient
one and the time-independent static one,
 $I(t) = I_{\mbox{stat}} + I_{\mbox{trans}}(t)$.
For the weak detuning we obtain\cite{PhysRevB.62.R16271,ourPRL}
\begin{eqnarray}
 I_{\mbox{stat}} = \frac{\Gamma}{4 \pi \Omega} \, \mbox{Im}
 \, \sum_{p=\pm} \, (\Omega +
 p \Gamma/2) \nonumber \\
 \times \Psi \left( \frac{1}{2} + \frac{i V + p \, \Omega +
 \Gamma/2}{2 \pi T} \right) \, ,
\end{eqnarray}
where $\Psi$ denotes the digamma function. The transient part is
given by
\begin{eqnarray}
 I_{\mbox{trans}}(t) &=& \frac{\Gamma T}{2 \Omega} \, \mbox{Im}
 \sum_{p= \pm}  \frac{( \Omega - p \, \Gamma/2) \, e^{(i V + p \, \Omega -
 \Gamma/2 - \pi T)t}}{i V + p \, \Omega - \Gamma/2 - \pi T} \,
 \nonumber \\ &\times&
\nonumber   {}_2{\bf F}_1\left( 1, \frac{1}{2} -
 \frac{i V + p \, \Omega - \Gamma/2}{2 \pi T};
 \right. \nonumber \\
  &&  \left. \frac{3}{2} -
 \frac{i V + p \, \Omega - \Gamma/2}{2 \pi T}; e^{ - 2 \pi T t}
 \right)  \, ,
\end{eqnarray}
where ${}_2{\bf F}_1$ denotes the hypergeometric function. For
weak detuning the transient current oscillates with the frequency
$\propto V$.\cite{PhysRevB.62.R16271} This situation changes in
case of strong detuning when $\Omega$ becomes imaginary and enters
the above equations in the same way as the voltage. Then there are
\emph{two} different frequencies $\propto |V\pm \Omega|$ and a
beating pattern emerges, s.
Fig.~\ref{MajoranaSwitchingCurrentPlots}. Of course these features
can be observed only when the oscillation period is smaller than
the competing time scale $\propto \Gamma + \pi T$, which governs
the overall current relaxation.

%\section{Calculation of the dot population/magnetization}

It turns out, that for the calculation of the dot population it is
more convenient to work with the lesser Keldysh GF
 $D_{ab}^<(t,t') = - i \langle a(t) \, b(t') \rangle$.
The calculation of this GF is accomplished using the Dyson
equation in the time domain (multiplication corresponds to time
integrations): $D_{ab} = D_{ab}^{(0)} + D_{ab}^{(0)} \, \Sigma \,
D_{bb}$, where the self-energy is proportional to the unperturbed
GF for the $\xi$-Majoranas $\Sigma = \Gamma g_{\xi \xi}$. After
the Keldysh disentanglement we obtain
\begin{eqnarray}                \label{longab}
 D_{ab}^< = D_{ab}^{(0) <} + D_{ab}^{(0) R} \,\Sigma^R \, D_{bb}^<
 \\ \nonumber
 + D_{ab}^{(0) R} \, \Sigma^< \, D_{bb}^A + D_{ab}^{(0) <} \, \Sigma^A \,
 D_{bb}^A  \, .
\end{eqnarray}
Thus the calculation of the inhomogeneous $D_{ab}$-GF is now
reduced to the calculation of the homogeneous $D_{bb}$. The
necessary zero order GFs are
\begin{eqnarray} \nonumber
 D^{(0) <}_{ab} (t) = - (\kappa/2) \, e^{ i \kappa \Delta t} \,
 , \, \,
 D_{ab}^{(0) R} (t) = - i \Theta(t) \, \sin (\Delta t) \, ,
\end{eqnarray}
where $\kappa=\pm 1$ encodes the initially populated/empty dot
level. For the retarded component of the homogeneous $D_{bb}$ one
obtains the equation
  $D_{bb}^R = D_{bb}^{(0) R} +  D_{bb}^{(0) R} \, \Sigma^R \,
  D_{bb}^R$.
\begin{figure}[t]
  \centering
  \vspace{0.3cm}
  \includegraphics[width = 0.40 \textwidth]{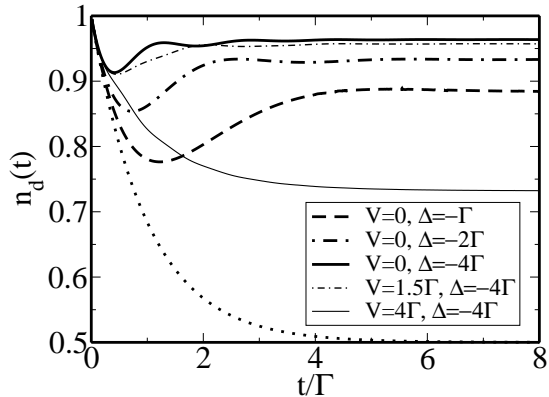}
  \caption{Population of the dot at zero temperature for
  different voltages $V$ and dot offset energies $\Delta$. The dotted
  line corresponds to the resonant case $\Delta=0$.}
  \label{PopulationPlots}
\end{figure}

The resonant case $\Delta=0$ is especially simple. Here the bare
retarded inhomogeneous GF vanishes and one obtains
\begin{eqnarray}                         \label{resonantcontrib}
 n_d (t) =  \left( 1 + \kappa e^{ - \Gamma t} \right)/2
 \, .
\end{eqnarray}
This is a very remarkable result since the relaxation of the dot
occupation/magnetization is independent on both temperature and
applied voltage and is only governed by the timescale $1/\Gamma$.

In the off-resonant case the calculations are more complex since
we need the full homogeneous $D^<_{bb}$. It is given
by\cite{Langreth}
%\cite{Langreth,mahan}
\begin{eqnarray}               \label{longformula} \nonumber
 D^<_{bb} = (1 + D^R_{bb} \, \Sigma^R ) \, D^{(0) <}_{bb} \, (1 + \Sigma^A \,
D^A_{bb})
%\\ \nonumber
 + D^R_{bb} \, \Sigma^< \, D^A_{bb} \, .
\end{eqnarray}
Putting the result into (\ref{longab}) we obtain for weak
detuning
%\footnote{We refrain from performing the energy
%integration as the emerging formulas are lengthy and less suitable
%for discussion.}
\begin{eqnarray} \nonumber
 n_d(t) =  \frac{1}{2} \left( 1 + \kappa e^{ - \Gamma t} \right) -
 \frac{\Gamma \Delta}{2 \Omega^2} \int \frac{d \omega}{2 \pi} \,
\frac{4 \Omega \, n_L}{(\Delta^2 - \omega^2)^2 + \omega^2
\Gamma^2}
\\ \nonumber
 \times \left\{ \Omega  \omega ( 1 + e^{-\Gamma
t})  -  e^{ -\Gamma t/2} \left[ (\omega^2 + \Delta^2) \sinh(\Omega
t) \sin(\omega t) \right. \right. \\ \nonumber \left. \left. + 2
\omega \Omega \cosh(\Omega t) \cos(\omega t)\right] \right\} \, .
\end{eqnarray}
The most striking feature of this result is that the relaxation
behavior of the information about the initial preparation is
exactly the same as in the resonant case and independent on either
the temperature or the applied voltage. Thus the rate at which the
system `forgets' its initial preparation does not depend on these
parameters. At zero temperature the effect of finite voltage is to
fix the upper energy integration boundary. Since the integrand is
an odd function of energy the whole correction due to finite
$\Delta$ vanishes towards larger $V$, so that $n_d$ approaches
that of a system at resonance, see Fig.~\ref{PopulationPlots}. The
effect of finite temperature is very similar: due to smearing off
of the Fermi edge the contribution of the energy integration
decreases so that the relative weight of the resonant contribution
(\ref{resonantcontrib}) increases and becomes more and more
dominant. For finite detuning $n_d$ oscillates with the frequency
$\propto \Delta$. As soon as the applied voltage becomes nonzero a
beating pattern emerges just as for the time dependence of the
transport current.

%\section{Conclusions}

%\acknowledgments

The author would like to thank T.~L.~Schmidt and L.~M\"uhlbacher
for many interesting discussions. The financial support was
provided by the DFG grant No.~KO~2235/2 and by the Kompetenznetz
``Funktionelle Nanostrukturen III'' of the Landesstiftung
Baden-W\"urttemberg (Germany).
\bibliography{MajoranaSwitchingPaper}

\end{document}